\documentclass[letter,traditabstract]{aa}
\usepackage{graphics}
\usepackage{txfonts}
\usepackage{amssymb}
\usepackage{lscape}
\usepackage{epsfig}
\usepackage{natbib}
\usepackage{multirow}

\usepackage{xcolor}

\begin{document}

\def\mpc{h^{-1} {\rm{Mpc}}} 
\def\kpc{h^{-1} {\rm{kpc}}}
\newcommand{\mincir}{\raise
-2.truept\hbox{\rlap{\hbox{$\sim$}}\raise5.truept\hbox{$<$}\ }}
\newcommand{\magcir}{\raise
-2.truept\hbox{\rlap{\hbox{$\sim$}}\raise5.truept\hbox{$>$}\ }}

\title{High density of active galactic nuclei in the outskirts of distant galaxy clusters} 

\author{E. Koulouridis \and I. Bartalucci}

\institute{AIM, CEA, CNRS, Universit\'e Paris-Saclay, Universit\'e Paris Diderot, Sorbonne Paris Cit\'e, F-91191 Gif-sur-Yvette, France}
\date{Received/Accepted}

\abstract{We present a study of the distribution of X-ray detected active galactic nuclei (AGN) in the five most massive, $M_{500}^{SZ}>10^{14} M_{\odot}$ , and distant, z$\sim$1, galaxy clusters in the \textit{Planck} and South Pole Telescope (SPT)\textit{} surveys. The spatial and thermodynamic individual properties of each cluster have been defined with unprecedented accuracy at this redshift using deep X-ray observations. This is an essential property of our sample in order to precisely determine the $R_{500}^{Y_{\textrm x}}$ radius of the clusters. For our purposes, we computed the X-ray point-like source surface density in 0.5$R_{500}^{Y_{\textrm x}}$ wide annuli up to a clustercentric distance of 4$R_{500}^{Y_{\textrm x}}$, statistically subtracting the background and accounting for the respective average density of optical galaxies. We found a significant excess of X-ray point sources between 2 and 2.5$R_{500}^{Y_{\textrm x}}$ at the 99.9\% confidence level. The results clearly display for the first time strong observational evidence of AGN triggering in the outskirts of high-redshift massive clusters with such a high statistical significance. We argue that the particular conditions at this distance from the cluster centre increase the galaxy merging rate, which is probably the dominant mechanism of AGN triggering in the outskirts of massive clusters.}

\authorrunning{E. Koulouridis \& I. Bartalucci}
\titlerunning{High density of active galactic nuclei in the outskirts of distant galaxy clusters}

\maketitle

\section{Introduction}

Supermassive black holes (SMBHs) are at the centre of modern astrophysical research today not only because they are hosted by every massive galaxy in the local Universe, but also because the evolution of the SMBH and its host galaxy appears tightly linked \citep[e.g.][]{Gultekin09,Zubovas12}. All SMBHs are thought to undergo active phases, the so-called active galactic nucleus (AGN) phases, during which they accrete the surrounding gas, and they emit an immense amount of energy. Theoretical models proposed that during this active phase, AGN produce a feedback wind that can explain the co-evolution of the SMBH and its host galaxy \citep[e.g.][]{Schawinski06,Cen11}. Therefore, the study of AGN is essential for understanding the cosmic history of accretion onto SMBHs and their relation to the host galaxy. However, we still do not fully understand the central engine, and the mechanisms that trigger or suppress AGN are still a topic of great debate. 

\begin{table*}[ht]
\caption{\footnotesize Sample of galaxy clusters}\label{tab:clus_prop}
\centering
    \begin{tabular}{lccccc}
\hline        
\hline
      Cluster            &  $z$        & $M_{500}^{SZ}$ $^{(b)}$ & \textit{Chandra} Obs. ID $^{(c)}$& Exposure $^{(d)}$\\
                         &                                 & [$10^{14}M_{\odot}$]           &                                  & [ks]         \\
\hline
      SPT-CLJ2146-4633   & $0.933$              &  $5.5 \pm 0.9$                 & \textbf{13469} & \textbf{71} \\
      \multirow{3}{105pt}{PLCKG266.6-27.3$^{(a)}$}& \multirow{3}{20pt}{$0.972$}   & \multirow{3}{30pt}{$8.5 \pm 0.7$} & \multirow{3}{150pt}{14017, \textbf{14018}, 14349, 14350, 14351, 14437, 15572, 15574, 15579, 15582, 15588, 15589} & \multirow{3}{40pt}{227, \textbf{37}}  \\ 
      &&&&\vspace{25pt}\\
      SPT-CLJ2341-5119   & $1.003$              & $5.6 \pm 0.9$                  & 9345, 11799 & 78, \textbf{50} \\
      SPT-CLJ0546-5345   & $1.066$                &  $5.1 \pm 0.8$                 & 9332, 10851, 10864, \textbf{9336}, 11739 & 68, \textbf{30} \\
      SPT-CLJ2106-5844   & $1.132$                & $7.0 \pm 0.4$                  & 12180, \textbf{12189} & 71, \textbf{53}\\
      \hline
    \end{tabular}
    
\tablefoot{$^{(a)}$ SPT name: SPT-CLJ0615-5746. $^{(b)}$ Masses published in the SPT catalogue \citep{bleem15spt}. $^{(c)}$ The Obs. ID in bold refers to the longest observation. $^{(d)}$ The exposure times reported in normal and bold refer to the merged datasets and the longest Obs. ID, respectively. The former and latter were used for point source detection and flux measurement, respectively.}
\end{table*}
There is compelling evidence that the presence of AGN is closely linked to the large-scale environment, and that galaxy mergers and interactions play an important role in AGN triggering and evolution \citep[e.g.][]{Koulouridis06,Hopkins08}. Clusters of galaxies are ideal laboratories for investigating the impact of dense environments on AGN demographics. As structures grow hierarchically, the majority of galaxies end up in clusters \citep[e.g.][]{Eke04,Calvi11}, which are therefore the predominant environment of galaxies and can play a very important role in establishing their properties. 
Previous studies have shown that AGN in clusters of galaxies are strongly affected by their environment, but in a complicated way. Their ability to accrete was found to depend on both their distance from the cluster centre and the mass of the cluster \citep[][]{Koulouridis18b,Ehlert15}. In more detail, the hot intra-cluster medium (ICM) is probably able to strip or evaporate the cold gas reservoir of galaxies \citep[e.g.][]{Gunn72,Cowie77,Giovanelli85,Popesso06a,Chung09,Jaffe15,Poggianti17b} and can strongly affect the fueling of the AGN. Several studies have indeed reported a significant lack of AGN in rich galaxy clusters with respect to the field \citep[][]{Haines12,Ehlert13,Ehlert14,Koulouridis10}. In addition, most of the X-ray sources found in clusters exhibit weaker optical AGN spectra than AGN in the field \citep[][]{Marziani17}, or show no signs of an optical AGN  \citep[e.g.][]{Martini06,Davis03}. However, \citet{Poggianti17a} suggested that ram pressure stripping may also act as a triggering mechanism for AGN activity in cluster members. In addition, there is evidence that the cluster mass also plays an important role in the efficiency of ram pressure stripping \citep[][]{Ehlert15,Koulouridis18b}, and furthermore, the fraction of AGN in cluster galaxies was reported to sharply increase with redshift \citep[][]{Martini13,Bufanda17}. Therefore, the physical mechanisms that enhance or suppress the AGN activity are still debated, especially at high redshift, where clusters become sparse and their properties are less well constrained.

In contrast to the lack of AGN in cluster centres, a number of studies have found a tentative excess of X-ray AGN in the outskirts \citep[][]{Ruderman05,Fassbender12,Haines12,Koulouridis14}, supporting the presence of an in-falling population, probably triggered by galaxy mergers  \citep[][]{Fassbender12,Ehlert15}. The excess was recently confirmed by a spectroscopic study \citep[][]{Koulouridis18b} of a homogeneous sample of 167 X-ray selected clusters of the XXL survey \citep[][]{Pierre16} up to $z=0.5$. However, it was correlated only with the less massive half of the sample ($M_{500}<10^{14}$ M$_{\odot}$), probably because of the high-velocity dispersions in massive clusters that may effectively reduce the galaxy merging rate \citep[see also][]{Arnold09}. Further studies are needed to investigate the excess and its cause, and to clarify how it is affected by cluster mass and redshift. If confirmed, it will add an important piece of information to our knowledge of AGN and their interplay with their local and large-scale environment. 

In this context, high-redshift and massive clusters are of particular interest. Firstly, they can be used to test the evolution by comparing their properties with local objects. Secondly, cosmological simulations \citep[][]{vazza11} showed that most massive clusters host more intense merging activities in the outskirts. Thus, they are ideal targets to study how the environment affects the AGN activity and its evolution. Unfortunately, the properties of these objects are poorly constrained at high redshifts because of the obvious observational limitations. Furthermore, they are intrinsically rare. Large-sky surveys based on the  the Sunyaev-Zel’dovich (SZ) effect \citep[][]{Sunyaev80} have been game-changers in this respect, the SZ being redshift-independent and characterised by a tight relation with the underlying halo mass. 
 In this study, we select the five most massive, $M_{500}^{SZ}> 5\times 10^{14}M_{\odot}$\footnote{$R_{\Delta}$ and $M_{\Delta}$ denote the radius at which the cluster density is $\Delta$ times the critical density of the Universe and the mass within, respectively. Throughout this work, we use $\Delta=500$.}, and distant, z$>$0.9, clusters detected in the SZ \textit{Planck }\citep[][]{plck_esz1,plck_psz1,plck_psz2} and South Pole Telescope \citep[SPT,][]{reichardt2013,bleem15spt} surveys. These objects have been investigated in detail by \citet[][]{bartalucci17,bartalucci18} using deep X-ray observations. X-ray observations are much less strongly affected by projection effects and allow accurate measurements of the cluster properties. They are also the most efficient way to detect AGN  \citep[][]{Brandt15}, which appear point-like and comprise the vast majority of sources.

In Sect. 2 of the paper we present the cluster and AGN samples, and in Sect. 3 we describe the applied method and results. Finally, we draw our conclusions and discuss the results in Sect. 5. Throughout this paper, we use $H_0=70$ km s$^{-1}$ Mpc$^{-1}$, $\Omega_m=0.3$, and $\Omega_{\Lambda}=0.7$.

\section{Sample selection and data analysis}

\subsection{Galaxy cluster sample.}

The sample contains the five most massive, $M_{500}^{SZ} \ge 5\times 10^{14}$ M$_{\odot}$, and distant, $z>0.9$, galaxy clusters in the SPT and \textit{Planck}   catalogues. All five objects have been observed by \textit{Chandra} using the Advanced CCD Imaging Camera \citep[][]{garmire03} and XMM-\textit{Newton} using the European Photon Imaging Camera \citep[][]{turner01,struder01}. In this work, we use the results of the analysis that characterised the spatial and thermodynamic properties of the five objects by combining the two instrument datasets \citep[][]{bartalucci17,bartalucci18}. In particular, we use the measurements of $R_{500}^{Y_{\textrm x}}$ reported in Table 1 of \cite{bartalucci18}.

\subsection{Point source analysis.}
\textit{Chandra} datasets were reprocessed and cleaned from flares using the Chandra Interactive Analysis of Observations \citep[][]{fruscione06} (CIAO) version $4.9$ and calibration database version $4.7.3$ \citep[][]{bartalucci17}. After the cleaning step, we merged multiple observations of the same object when available to maximise the statistic. We ran the CIAO \textit{wavdetect} tool \citep[][]{freeman02} on exposure-corrected images in the [0.5-2], [2-8], and [0.5-8] keV bands and merged the resulting catalogues. We inspected the merged catalogue for missed or false detections by eye. We measured each point source flux with the following scheme: (i) the first estimate was obtained using the CIAO \textit{srcflux} tool folding an absorbed power law to model the point source emission. The column density absorption was accounted for using the WABS model \citep[][]{morrison83} fixed to the LAB survey \citep[][]{kalberla05}, and the index of the power law was fixed to 1.7. We approximated the point spread function (PSF) using the \textit{arfcorr} tool, which generates a circular region of radius R encircling $90\%$ of the flux, and an external annulus, [1-5]R, to measure the background. (ii) We used the flux estimate to simulate the \textit{Chandra} response to a point source emission with the MARX tool \citep[][]{davis12}, version $5.3.2$. We obtained the precise shape of the PSF by fitting an ellipse encircling $90\%$ of the flux to the simulated image. The ellipsoidal annulus between one and five times the PSF ellipse semi-axes was used to extract the background. We then measured the point source flux repeating step (i) and folding the refined PSF and background region. The flux limits in our five regions vary between $8.9\times10^{-16}$ to $3\times10^{-15}$ erg s$^{-1}$ cm${-2}$ for the [0.5-2] keV band and $2.9\times10^{-15}$ to $9.9\times10^{-15}$ erg s$^{-1}$ cm${-2}$ for the [2-8] keV band.
It is worth noting that we used the longest single observation for each cluster, reported in Table 1, to measure the flux to avoid cross-calibration problems between observations with different settings, that is, the same point source is observed with different parts of the detector.

To estimate the expected number of X-ray point sources in the field, first we produced flux-limit maps and the effective area curve of each cluster region following the recipe of \citet{branchesi2007} (Appendix C), folding both instrumental and background effects. The area curve (cumulative) determines at each flux $S$ the maximum area $\Omega$ of the X-ray observations where a point-like source of this flux can be detected. Then, we selected the cumulative flux distribution (the so-called logN-logS relation), which was computed by \citet{Moretti03} from the combined data of six different surveys. The log$N$-log$S$ relation defines at each flux $S$ the number of all sources $N(S)$ brighter than S, weighted by the corresponding sky coverage. Finally, folding the effective area curve of each cluster region in the log$N$-log$S$ relation, we derived the respective field surface density of X-ray point-like sources.

\section{Method and results}

\begin{figure}
\centering
\resizebox{8cm}{12cm}{\includegraphics{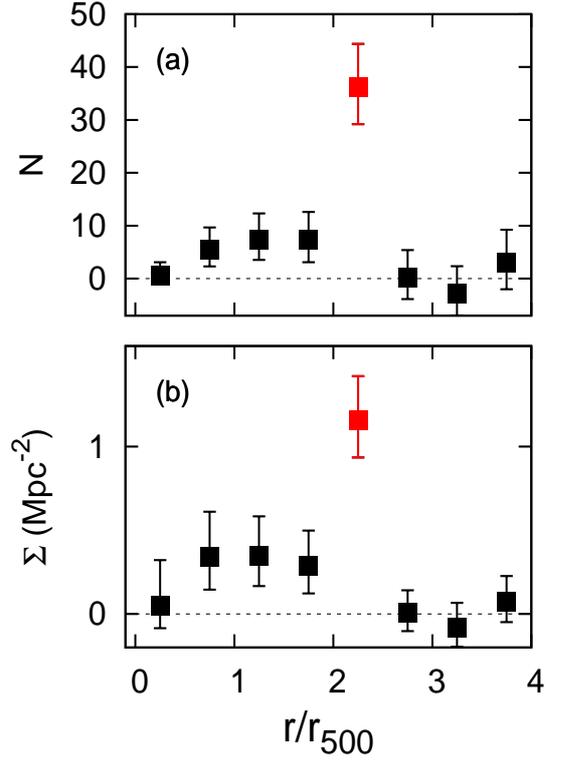}}
\caption{Top: Distribution of X-ray point sources in galaxy clusters as a function of radius in units of $R_{500}^{Y_{\textrm x}}$. Bottom: Total surface density of X-ray point sources in galaxy clusters divided by the optical galaxy profile. In both panels, the expected number of sources in the field has been subtracted statistically from each annulus. Error bars indicate the 1$\sigma$ confidence limits for small numbers of events in astrophysical data. A significant excess is found in the cluster outskirts between 2$R_{500}^{Y_{\textrm x}}$ and 2.5$R_{500}^{Y_{\textrm x}}$ at the 99.9\% confidence level.}
\end{figure}

\begin{figure}
\centering
\resizebox{8cm}{10cm}{\includegraphics{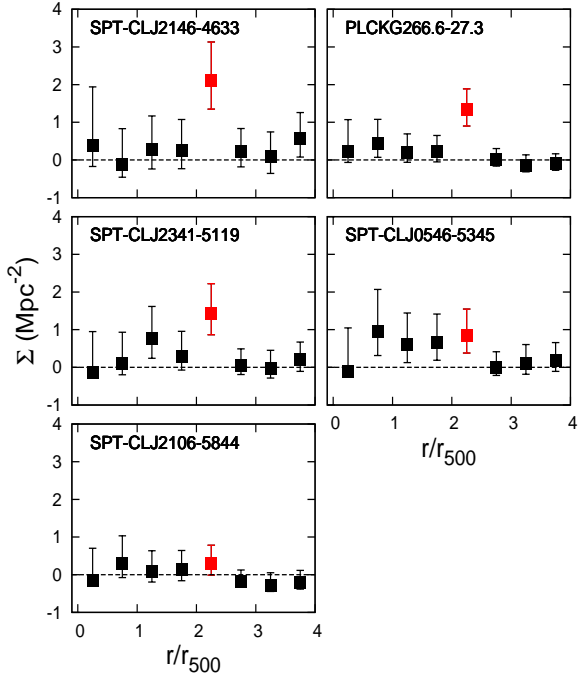}}
\caption{Surface density of X-ray point sources in galaxy clusters divided by the optical galaxy profile as a function of radius. Error bars indicate the 1$\sigma$ confidence limits for small numbers of events in astrophysical data. Except for the bottom panel, a significant X-ray point source excess is found between 2$R_{500}^{Y_{\textrm x}}$ and 2.5$R_{500}^{Y_{\textrm x}}$ at the 99.9\% confidence level (99\% for SPT-CL0546-5345).}
\end{figure}
To investigate the effect of the cluster environment on AGN activity, we first sampled the X-ray point sources detected above 3$\sigma$ up to $4R_{500}^{Y_{\textrm x}}$. We divided the area into eight $\frac{1}{2}R_{500}^{Y_{\textrm x}}$ concentric annuli centred on the X-ray peak emission of the cluster, determined in $[0.5-2.5]$ keV \textit{Chandra} count-rate images. We then computed the number of X-ray point sources, $X_i$, in each annulus $i=1,2,...,8$. We accounted for incomplete coverage of the last two annuli by introducing in each case the appropriate weight in the X-ray number counts. PLCKG266.6-27.3 and SPT-CLJ2106-5844 present the highest incompleteness, where almost one-third of the eighth and a small part of the seventh annulus are not covered by the detector. The effect of using a different centre for the cluster analysis has been investigated in \cite{bartalucci17} using the position of the brightest cluster galaxy (BCG) as centre. The results were consistent with the analysis performed using the X-ray peak.

The corresponding number of X-ray point sources in the field, $F_i$, was derived by folding the effective area curves of the cluster regions to the log$N$-log$S$ relation (see section 2.2). All X-ray-detected point-like sources are potentially AGNs if located at the redshift of the cluster because their X-ray luminosity would exceed $L_{\rm X [0.5-8]\,keV}> 3\times10^{42}$ erg s$^{-1}$.

In Fig. 1(a) we plot the number of X-ray sources found in excess of the field value, $N=X-F$, as a function of distance from the cluster centre. We clearly detect a significant excess of X-ray point sources in the outskirts of our clusters (marked in red in the plot). However, the density of X-ray point sources in each annulus should further be compared with the corresponding density of cluster galaxies, which is expected to increase sharply towards the cluster centre. For this purpose, we calculated a weighting factor $w_i=G_i/G_f$, where $G_i$ is the surface density of optical galaxies in each annulus $i$, and $G_f$ is the respective density in the field. The values of $G_i$ and $G_f$ were derived from the average optical profile computed in \citet[][]{Ehlert14} of a population of 42 clusters between $z=0.2-0.7$, with a mass range similar to our sample. This is a reasonable approximation when we assume self-similar evolution. The optical profile covers up to 2.5$R_{500}$ radius, where they still find an excess of optical galaxies compared to the field. However, we assumed that above 3$R_{500}$ , it reaches the background, and we interpolated only for the sixth annulus. We note that this empirical profile is consistent with a Navarro-Frenk-White profile \citep{Navarro97}, which was shown to describe the distribution of galaxies in SPT clusters well \citep{Hennig17}. Therefore, for each annulus $i$ we computed the sum of all five clusters $j$ as follows:
\begin{equation}
\Sigma_i=\sum_{j=1}^5(X_{ij}-F_{ij})/w_i A_{ij},
\end{equation}
where $A_{ij}$ is the area in Mpc$^2$ of each annulus $i$ in every cluster $j$.
In Fig. 1(b) we plot the surface density $\Sigma$ as a function of the distance from the cluster centre. The excess of X-ray point sources in the outskirts is statistically significant at the 99.9\% confidence level (we here always use the confidence level for small numbers of events in astrophysical data \citep[][]{Gehrels86}). This strongly indicates an excess of AGN in our clusters, correlated with the galaxy population in their outskirts, between 2 and 2.5$R_{500}^{Y_{\textrm x}}$. The respective results obtained using different luminosity thresholds for the X-ray point sources are presented in Fig. A.1 in the appendix.

To clarify if the detected excess is a general property of our sample or is due to a large number of sources in just one or two clusters, we plot in Fig. 2 the surface density results for each cluster separately (for the X-ray point source distribution, see Fig. A.2). In each panel, the fifth annulus is marked in red. It corresponds to the overdensity found between $2R_{500}^{Y_{\textrm x}}$ and 2.5$R_{500}^{Y_{\textrm x}}$ in Fig. 1. In clusters SPT-CLJ2146-4633, PLCKG266.6-27.3 and SPT-CLJ2341-5119, we clearly detect a visible excess that is statistically significant at the 99.9\% confidence level. In the case of SPT-CLJ0546-5345, the corresponding confidence level is 99\%, and in addition, a more extended excess of X-ray point sources is present between $0.5R_{500}^{Y_{\textrm x}}$ and 2.5$R_{500}^{Y_{\textrm x}}$. As the only exception, in SPT-CLJ2106-5844, the cluster with the highest redshift, the excess in the fifth annulus is not statistically significant. In Fig. 3 we present as an example the X-ray image of PLCKG266.6-27.3, where we overplotted the large number of point-source detections in the fifth annulus.

\begin{figure}
\centering
\includegraphics[scale=0.35]{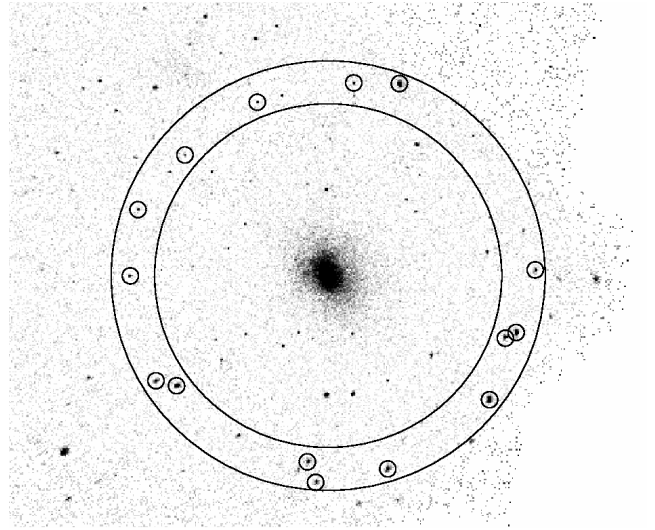}
\caption{X-ray image (0.5--2.0 keV) of cluster PLCKG266.6-27.3 at $z=0.972$, with an estimated mass of 8.5$\times 10^{14}$ M$_\odot$. Small circles mark the detected point sources in the fifth annulus, where a large excess of X-ray sources with respect to the field was found. Large circles mark the boundaries of the fifth annulus, 2 - 2.5$R_{500}^{Y_{\textrm x}}$.}
\end{figure}

\section{Conclusions and discussion}

We studied the AGN activity in massive and distant clusters as a function of clustercentric distance. To this end, we have used a sample of five clusters at $z\sim1$ with uniquely well-defined properties for this redshift, which allowed us to accurately determine their $R_{500}^{Y_{\textrm x}}$ radius and to divide the cluster regions into fine bins. Our results showed a highly significant excess of X-ray point sources between 2 and 2.5$R_{500}^{Y_{\textrm x}}$, which strongly suggests a high occurrence of AGN triggering in the cluster outskirts. This distance is in agreement with the results of \citet[][]{Ruderman05}, who reported a mild excess of X-ray sources between 1.5 and 3 Mpc in 24 dynamically relaxed massive clusters spanning the redshift range $z=0.3-0.7$. However, no excess was found in that work in the disturbed clusters, although the dynamical state classification is rudimentary, as the authors quote. In contrast, with the exception of PLCKG266.6-27.3, our clusters are disturbed based on a very thorough analysis of their dynamical status \citep[][]{bartalucci17}. More recently, \citet{Fassbender12} also showed a similar excess of X-ray sources between 4 and 6 arcmin (2-3Mpc) from the centres of 22 massive clusters spanning the redshift range $z=0.9-1.6$. They argued that at this distance, the combination of still low relative galaxy velocities and already high source density can increase the merging rate, which can lead to AGN triggering. We note, however, that no similar excess was found in a recent study of optical, infrared, and radio AGN in 2300 infrared-selected clusters \citep{Mo18}.

Theoretically, non-axisymmetric perturbations can cause mass inflow during galaxy interactions and merging, and can lead to AGN triggering \citep[][]{Koulouridis13,Koulouridis14a,Ellison11,Villforth12,Hopkins14}. Therefore,
the detected AGN excess can be explained by a high rate of galaxy merging \citep[e.g.][]{Ehlert15} caused by the particular conditions in the cluster outskirts. In more detail, according to the cold dark matter (CDM) paradigm of hierarchical structure formation, many galaxies experience high-density environments before they become cluster members, either as members of smaller groups or by forming within large-scale filaments. \citet{McGee09} studied a simulated galaxy cluster and group catalogue drawn from the Millennium Simulation \citep{Springel05} and found that up to redshift 1.5, galaxy clusters have accreted a significant fraction of their final galaxy populations through galaxy groups. In a similar study, \citet{DeLucia12} argued that a large portion of cluster galaxies could have been subject to pre-processing in group environments. Most importantly, \citet[e.g.][]{Vijayaraghavan13} studied a merger between a group and a cluster in an $N$-body cosmological simulation and then performed an idealised hydrodynamical simulation of the merger. Interestingly, they showed that the merging rate of the infalling group galaxies steadily increases until the first pericentric passage. However, so does the ram pressure, which can strip the gas from the galaxy and have the opposite effect on AGN fueling. It is possible that our results pinpoint a specific location where the infalling galaxy density is already significantly enhanced \citep{Fassbender12} and the merging rate becomes high enough to trigger a large number of AGN before ram pressure stripping can effectively hamper their fueling capability.  

A small excess of point-like sources is also detected in our cluster sample within 2$R_{500}^{Y_{\textrm x}}$ radius. This is in general agreement with previous studies, which reported that the AGN density in clusters above $z=1$ is at least equal to or higher than the respective AGN density in the field \citep[][]{Fassbender12,Martini13,Bufanda17}. Low-mass protoclusters at higher redshifts may contain even more AGN \citep[][]{Lehmer13,Krishnan17}. These findings are in sharp contrast with the strong suppression of AGN activity at low redshifts and support an evolution of the AGN fraction in cluster galaxies. We note that a dependence of the AGN density on cluster mass was also reported \citep[][]{Koulouridis14,Koulouridis18b,Ehlert15}, rendering the interpretation of the phenomenon even more complicated.
  
Our results provide observational evidence of the physical mechanisms that drive AGN and galaxy evolution within clusters, testing the efficacy of galaxy merging and ram pressure stripping within dense environments. The novelty of this work lies in the unique selection of massive clusters at $z\sim1$ and the unprecedented accuracy with which the physical properties of these clusters were measured at this high redshift. The AGN excess in the cluster outskirts peaks within a few hundred kpc, while the extent and the distance from the center of this region varies from cluster to cluster, depending on their physical properties. We argue that applying any binning method that does not account for individual cluster properties when stacking AGN number counts, for instance, fixed projected radius, fixed physical radius, or no binning, would dilute the excess. Consequently, future investigations should not only seek to maximise the samples, but also to better determine the cluster properties. Our future plan is to obtain follow-up optical observations of the X-ray AGN hosts in order to confirm their redshift and determine their properties. 

\begin{acknowledgements}
The results reported in this article are based on data obtained from the Chandra Data Archive.  
EK and IB thank CNES and CNRS for support of post-doctoral research. IB acknowledges the funding from the European Research Council under the European Union’s Seventh Framework Programme (FP72007- 2013) ERC grant agreement no 340519.
\end{acknowledgements}

\bibliographystyle{aa} 
\bibliography{mylib}

\begin{appendix}
\newpage
\section{Supplementary information}

\begin{figure*}[t]
\centering
\includegraphics{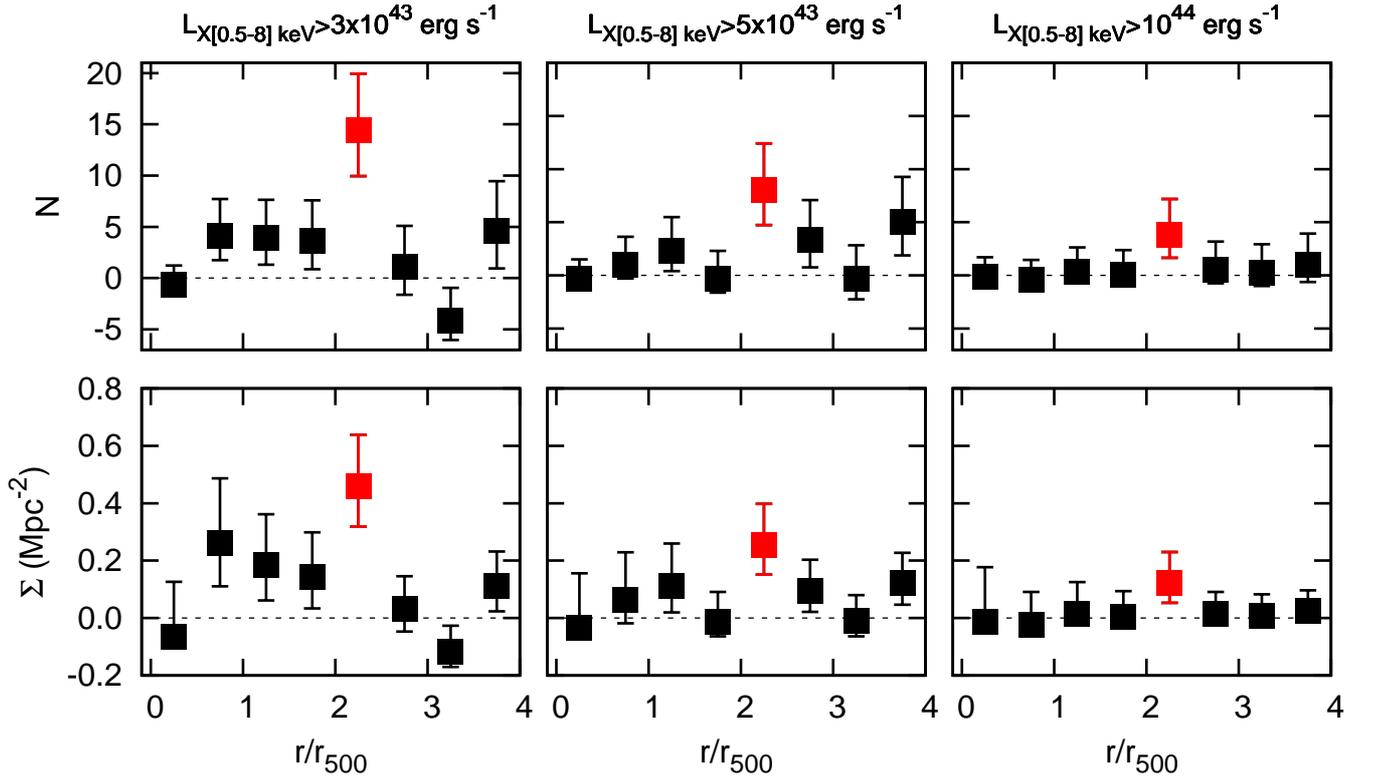}
\caption{Results obtained using increasing luminosity thresholds for the X-ray point sources. Top: Distribution of X-ray point sources in galaxy clusters as a function of radius in units of $R_{500}^{Y_{\textrm x}}$. Bottom: Total surface density of X-ray point sources in galaxy clusters divided by the optical galaxy profile. In both panels, the expected number of sources in the field has been subtracted statistically from each annulus. Error bars indicate the 1$\sigma$ confidence limits for small numbers of events in astrophysical data.}
\end{figure*}

To investigate if there is any correlation between the X-ray luminosity of the AGNs and the excess in the outskirts, we plot in Fig. A.1 the results of our analysis using an increasing luminosity threshold. The excess is present in all panels, even in the case of $L_{\rm X [0.5-8]\,keV}>10^{44}$ erg s$^{-1}$, where only a small number of point sources is detected above this limit in our clusters. 

In Fig. A.2 we present the distribution of X-ray point sources in the five individual clusters of our sample.

\begin{figure}[b]
\centering
\resizebox{8cm}{10cm}{\includegraphics{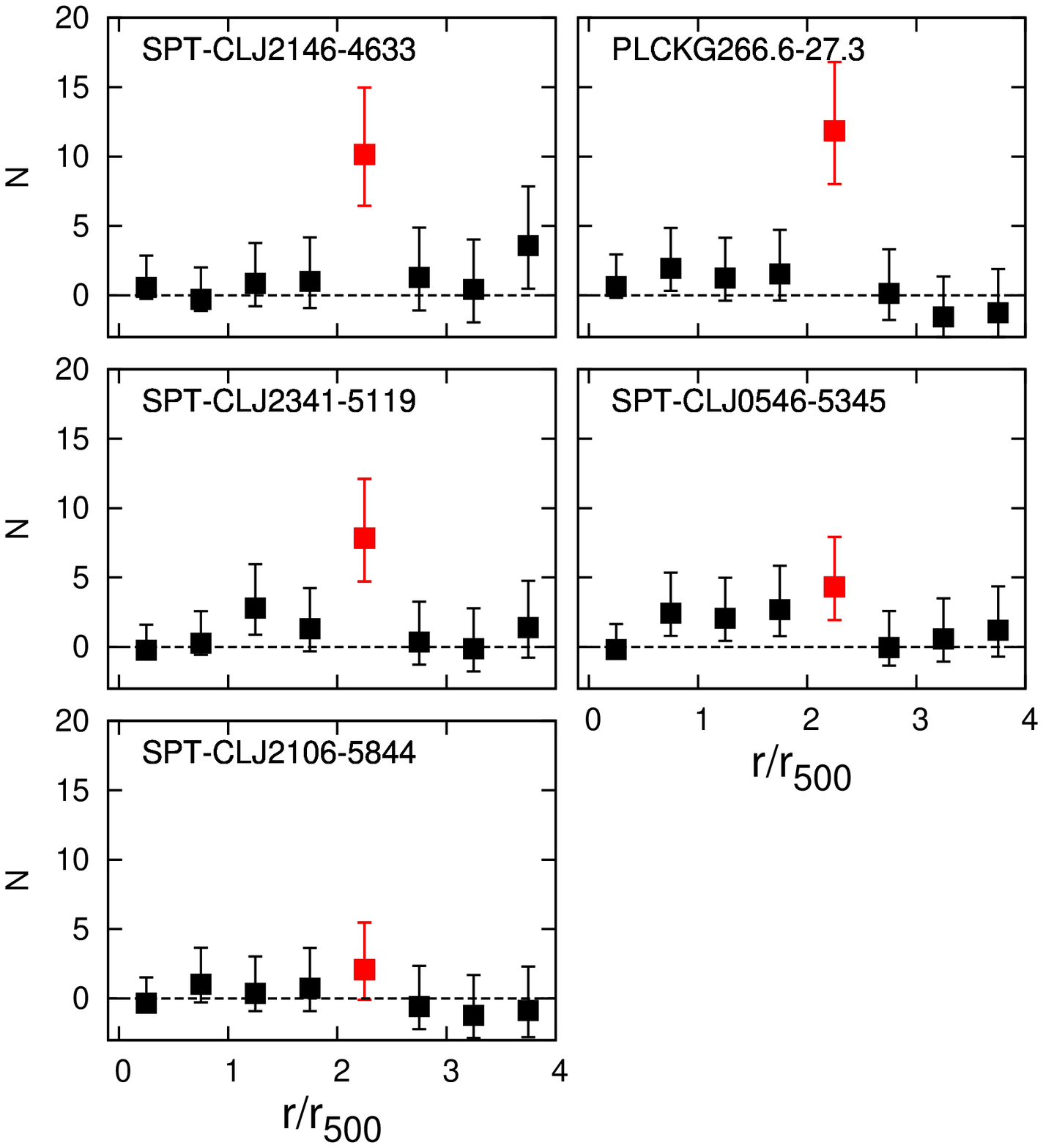}}
\caption{Distribution of X-ray point sources in galaxy clusters as a function of radius in units of $R_{500}^{Y_{\textrm x}}$. The expected number of sources in the field has been subtracted statistically from each annulus. Error bars indicate the 1$\sigma$ confidence limits for small numbers of events in astrophysical data.}
\end{figure}

\end{appendix}

\end{document}